\documentclass[aps,pre,groupedaddress,preprint]{revtex4-1}
\usepackage{amsmath}
\usepackage{physics}
\usepackage{graphicx}
\usepackage{subfigure}
\usepackage{hyperref}
\begin{document}
\title{Kinetic theory of $\sech^2x$ electron holes and applications to Kappa-distributed plasmas} 
\author{Ran Guo}
\email{E-mail: rguo@cauc.edu.cn}
\affiliation{Department of Physics, College of Science, Civil Aviation University of China, Tianjin 300300, China}
\begin{abstract}
    The kinetic theory of $\sech^2 x$-type electron holes is studied.
    The potential of the electron holes is solved in the weak amplitude limit by the pseudo-potential method. 
    We investigate the existence condition of the $\sech^2 x$ electron holes.
    It indicates that the derivatives of trapped and untrapped distributions at the separatrix play significant roles in determining the potential profile.
    The theory is then applied to the Kappa-distributed plasmas.
    The amplitude and width of the $\sech^2 x$ electron holes are analyzed.
    Finally, the theoretical results are verified by numerical calculations. 
\end{abstract}
\maketitle

\section{Introduction}
\label{sec:intro}
Electron hole (EH) in the phase space is an interesting nonlinear structure in plasmas.
It was first observed in the simulations of two-stream instabilities by Morse \textit{et al.} \cite{Morse1969}. 
After that, more and more investigations found that the EHs could form in observations and laboratory experiments \cite{Lefebvre2010,Fox2012,Hutchinson2018a,Liu2019,Mozer2021}. 
Liu \textit{et al.} identified the drifting EHs during the geomagnetically quiet time from observational data \cite{Liu2019}. 
Mozer \textit{et al.} observed EHs, as well as ion holes, on the Parker Solar Probe \cite{Mozer2021}.
They found that these EHs were probably produced by the nonlinear electron-streaming instabilities.
In addition, the EHs also attract theoretical interest to analyze their physical properties.
Vasko \textit{et al.} studied the EHs propagating in a weakly inhomogeneous magnetic field \cite{Vasko2016}. 
Hutchinson and Zhou investigated the various kinematic properties of EHs from the theories and simulations \cite{Hutchinson2016,Zhou2016}.
Aravindakshan \textit{et al.} studied the Gaussian type EHs by the Bernstein-Greene-Kruskal (BGK) integral approach in the background of thermal and superthermal space plasmas \cite{Aravindakshan2018}.

Among these theoretical works, an important branch is the kinetic description of EHs.
This theory was firstly given in the well-known paper of Bernstein, Greene, and Kruskal \cite{Bernstein1957}.
Then, Schamel developed the theory by assuming a specific distribution (hereafter Schamel distribution) \cite{Schamel1972,Korn1996,Schamel2015,Schamel2020}
and derived different solitary wave solutions in small amplitude limits, including $\sech^4 x$-type solution, $\sech^2 x$-type solution, and Gaussian type solutions \cite{Schamel1972,Korn1996,Schamel2020a}.
Turikov obtained the trapped-particle distribution of $\sech^{\nu} x$-type solitons, where $\nu$ is an arbitrary positive number \cite{Turikov1984}.
Goldman \textit{et al.} proposed a general theory of $\sech^4 x$ EHs, which interpreted the weak bipolar fields observed in space \cite{Goldman2007}.
Recently, Haas generalized the Schamel theory of $\sech^4 x$ EHs to the non-Maxwellian plasmas with singularities \cite{Haas2021}.
However, the kinetic theory investigations of $\sech^2 x$ EHs are rare.
Korn and Schamel showed that \cite{Korn1996} the solitary wave solution was of the $\sech^2 x$ form when the Schamel distribution for the trapped particles satisfied the specific condition $\beta=1-u^2$, where $\beta$ is the trapping parameter, and $u$ is the EH speed.
% More details can be found below Eq. \eqref{eq:ft_M} in Sec. \ref{sec:app_theo}.
To our knowledge, the general existence conditions of $\sech^2 x$ EHs have not been investigated through the kinetic theory.
Such potential profiles are always studied as a solitary wave solution of the Korteweg-de Vries equation \cite{Kourakis2012,Ali2017}.

In the studies of EHs, the distribution of untrapped particles plays an important role.
One always assumes the shifted-Maxwellian distribution for these free particles in the literature \cite{Schamel1972,Korn1996,Schamel2015,Schamel2020,Hutchinson2017,Schamel2020a}.
However, more and more works indicate that the plasmas are not always in equilibrium.
It shows that the Kappa distribution is more suitable than the Maxwellian one to model the non-equilibrium plasmas 
in space observations \cite{Schippers2008,Pierrard2016}, laboratory experiments \cite{Hellberg2000}, and computer simulations \cite{Vocks2008,Randol2014,Tao2014,Guo2020}.
The Kappa distribution has been widely applied to study various physical phenomena in plasmas, for instance, electron firehose instabilities \cite{Lopez2019,Nazeer2021}, transport coefficients \cite{Wang2017,Wang2018,Guo2019}, whistler instabilities \cite{Lazar2019,Abdul2021}, and electron acoustic waves \cite{Baluku2011,Guo2021a}.

The purpose of this work is to investigate the general kinetic theory of $\sech^2 x$ EHs in weak amplitude limits and to study the physical properties of such EHs in Kappa-distributed plasmas.
To achieve the above goal, we organize the paper as follows.
In Sec. \ref{sec:theo}, the general theory of $\sech^2 x$ EHs is given by the pseudo-potential method.
Then, the theory is applied to the Kappa-distributed plasmas in Sec. \ref{sec:app_theo}.
The parameter spaces of the potential amplitude and width are discussed.
% The physical behaviors of $\sech^2 x$ EHs are studied in a special case that the trapped and untrapped distributions are smooth at the separatrix in Sec. \ref{sec:app_sp}.
Numerical solutions are obtained and compared with the theoretical results for verifications in Sec. \ref{sec:app_num}.
At last, we conclude our results in Sec. \ref{sec:sum}.

\section{General theory}
\label{sec:theo}
We consider a one-dimensional (1D) electrostatic plasma.
For convenience, the dimensionless parameters are used in this study.
The length is scaled by the Debye length $\lambda_D = \sqrt{\epsilon_0 k_B T/(n_0 e^2)}$ and the velocity by the thermal speed $\sqrt{k_BT/m}$.
The potential is measured in the unit of $k_BT/e$ and the energy in the unit of $k_BT$.
The number densities of electrons and ions are scaled by $n_0$.
In the above notations, $\epsilon_0$ is the vacuum permittivity, 
and $e$ is the elementary charge.
$T$ and $n_0$ are, respectively, the kinetic temperature and the number density of undisturbed electrons at $x\rightarrow \pm \infty$. 

If a solitary wave propagates in the plasmas,
the potential in the wave frame is a stationary localized structure.
The EH indicates that the potential is positive everywhere, so the trapped species is the electron.
We suppose that the potential $\phi(x)$ goes to zero when $x \rightarrow \pm \infty$ and has a positive maximum $\psi$ at $x=0$.
In such a system,
the ions are assumed to be immobile and spatially uniform.
The electrons follow the stationary Vlasov-Poisson equations.
As we know, the 1D electron distribution solved from the stationary Vlasov equation must be a function of energy $W=v^2/2-\phi$.
Due to the non-uniform potential, 
the electrons are divided into the trapped electrons for $W<0$, and the untrapped electrons for $W\ge0$.
We denote the distributions of these two electrons as $f_t(W)$ and $f_u(W)$, respectively.
For the untrapped electrons, the left and right passing electrons may have different distributions.
They are represented by $f_u^+$ for $v>\sqrt{2\phi}$ and $f_u^-$ for $v<-\sqrt{2\phi}$.
The potential $\phi(x)$ is governed by the Poisson equation,
\begin{equation}
    \dv[2]{\phi}{x} = n-1,
    \label{eq:Poisson}
\end{equation}
where the dimensionless ion number density is set as $1$. The electron number density $n$ is given by,
\begin{equation}
    n = \int_{-\infty}^{-\sqrt{2\phi}} f^-_u \dd{v} + \int_{-\sqrt{2\phi}}^{\sqrt{2\phi}} f_t \dd{v} + \int_{\sqrt{2\phi}}^{+\infty} f^+_u \dd{v}.
    \label{eq:n_phi}
\end{equation}
The above Eq. \eqref{eq:n_phi} can be rewritten as,
\begin{equation}
    n = I_1+I_2, 
    \label{eq:n_phi_rewritten}
\end{equation}
where
\begin{equation}
   I_1 = \int_{-\infty}^{+\infty} f_u \dd{v},
   \label{eq:n_phi_i1}  
\end{equation}
and
\begin{equation}
   I_2 = \int_{-\sqrt{2\phi}}^{\sqrt{2\phi}} f_t \dd{v} - \int_{-\sqrt{2\phi}}^{\sqrt{2\phi}} f_u \dd{v},
   \label{eq:n_phi_i2}
\end{equation}
In the above equations, the domain of $f_u^+$ is analytically continued from $(\sqrt{2\phi},+\infty)$ to $(0,+\infty)$, and $f_u^-$ from $(-\infty,-\sqrt{2\phi})$ to $(-\infty,0)$.
We let $f_u^\pm$ have the identical functional forms in the original and continued domains.
It should be noticed that there are no untrapped electrons with the speed $\abs{v}<\sqrt{2\phi}$.
The above analytic continuation is just a kind of mathematical treatment.
In the small amplitude limit $\phi<\psi \ll 1$, we can expand $I_1$ into a power series,
\begin{equation}
    I_1 =1+\sum_{k=1}^\infty \left[\frac{(-1)^k(2k-1)!!}{k!} \pv{\int_{-\infty}^{+\infty} \frac{F_u}{v^{2k}} \dd{v}}\right] \cdot \phi^k, 
    \label{eq:I1_series}
\end{equation}
where $F_u=f_u|_{\phi=0}$ is the free electron distribution at $x\rightarrow \pm \infty$ and $\pv$ denotes the Cauchy principal value of the integral.
The detailed derivations are shown in Appendix \ref{sec:dev_I1}.
For the integral $I_2$, we introduce $\xi = \sqrt{-2W}=\sqrt{2\phi-v^2}$ which is much less than $1$ in the range of the integrations $(-\sqrt{2\phi},\sqrt{2\phi})$.
Consequently, we can expand $I_2$ at $\xi = 0$ and integrate $\int_{-\sqrt{2\phi}}^{\sqrt{2\phi}} \xi^k \dd{v}$,
\begin{align}
   I_2 = \sum_{k=0}^\infty\left[2^{\frac{k+1}{2}}\frac{\sqrt{\pi}\Gamma\left(\frac{k+2}{2}\right)}{k!\Gamma\left(\frac{k+3}{2}\right)} \eval{\dv[k]{}{\xi}\left(f_t - \frac{f_u^++f_u^-}{2} \right) }_{\xi=0} \right] \cdot \phi^{\frac{k+1}{2}}.
   \label{eq:I2_series}
\end{align}
Substituting Eqs. \eqref{eq:I1_series} and \eqref{eq:I2_series} into Eq. \eqref{eq:n_phi_rewritten}, one can rearrange the series expansion of $n$ as,
\begin{equation}
    n(\phi) = 1 +\sum_{k=0}^\infty \left( a_k \phi^{k+\frac{1}{2}}+b_k\phi^{k+1} \right),
   \label{eq:n_series}
\end{equation}
where the expansion coefficients $a_k$ and $b_k$ are, respectively,
\begin{equation}
    a_k = \frac{2^{k+\frac{3}{2}}}{(2k+1)!!(2k-1)!!} \eval{\dv[2k]{}{\xi}\left(f_t - \frac{f_u^++f_u^-}{2} \right) }_{\xi=0},
    \label{eq:ak}
\end{equation}
and
\begin{equation}
    b_k = \frac{\pi}{(2k)!!(k+1)!} \eval{\dv[2k+1]{}{\xi}\left(f_t - \frac{f_u^++f_u^-}{2} \right) }_{\xi=0}
               +\frac{(-1)^{k+1}(2k+1)!!}{(k+1)!} \pv{\int_{-\infty}^{+\infty} \frac{F_u}{v^{2k+2}} \dd{v}}.
    \label{eq:bk} 
\end{equation}
The above approach is similar to the method in Ref. \cite{Korn1996}, but we derive the general formulas for the expansion coefficients.
We neglect the higher-order terms $O(\phi^\frac{5}{2})$ in Eq. \eqref{eq:n_series},
\begin{equation}
    n(\phi) = 1 + A \phi^\frac{1}{2} + B \phi + C \phi^\frac{3}{2} + D \phi^2,
    \label{eq:n_abcd}
\end{equation}
where
\begin{equation}
    A = 2\sqrt{2} \eval{\left(f_t-\frac{f_u^++f_u^-}{2}\right)}_{\xi=0},
    \label{eq:A}
\end{equation}
\begin{equation}
    B = \pi \eval{\dv[]{}{\xi}\left(f_t-\frac{f_u^++f_u^-}{2}\right)}_{\xi=0}-\pv{\int_{-\infty}^{+\infty} \frac{F_u}{v^2} \dd{v}},
    \label{eq:B}
\end{equation}
\begin{equation}
    C = \frac{4\sqrt{2}}{3} \eval{\dv[2]{}{\xi}\left(f_t-\frac{f_u^++f_u^-}{2}\right)}_{\xi=0},
    \label{eq:C}
\end{equation}
\begin{equation}
    D = \frac{\pi}{4} \eval{\dv[3]{}{\xi}\left(f_t-\frac{f_u^++f_u^-}{2}\right)}_{\xi=0}+\frac{3}{2}\pv{\int_{-\infty}^{+\infty} \frac{F_u}{v^4} \dd{v}},
    \label{eq:D}
\end{equation}

The profile of the potential is determined by these expansion coefficients $A,B,C,D,\dots$.
We assume the distribution is continuous at the separatrix, i.e., $f_t=f_u^+=f_u^-$ at $W=0$, resulting in $A=0$.
Further, if $B,C \neq 0$, then, neglecting $O(\phi^2)$, one derives the potential $\phi(x) \propto \sech^4(\sqrt{B}x/4)$ well-known in the literature \cite{Schamel1972,Goldman2007}. 

However, according to the pseudo-potential method \cite{Hutchinson2017}, if, 
\begin{equation}
    A,C=0 \quad \text{and} \quad B,D \neq 0,
    \label{eq:sech2_cond}
\end{equation}
then, neglecting the higher-order terms $O(\phi^{5/2})$, we can derive the potential
\begin{equation}
    \phi(x) = \psi \sech^2 \left(\frac{x}{\Delta}\right),
    \label{eq:phi}
\end{equation}
with the amplitude
\begin{equation}
    \psi = -\frac{3B}{2D},
    \label{eq:psi}
\end{equation}
and the width
\begin{equation}
    \Delta = \frac{2}{\sqrt{B}}.
    \label{eq:delta}
\end{equation}
The detailed derivations could be found in Appendix \ref{sec:dev_phi}.
It indicates that the potential must be of the form \eqref{eq:phi} for arbitrary distributions satisfying the criteria \eqref{eq:sech2_cond} in the weak amplitude limit.
The different distributions determine the amplitude and the width of the potential through the coefficients $B$ \eqref{eq:B} and $D$ \eqref{eq:D}.

In Ref. \cite{Goldman2007}, Goldman \textit{et al.} proved that the EH potential should be the unique $\sech^4 x$ form in weak amplitude limit if the energy derivatives of $f_t$ and $f_u$ are not equal at the separatrix.
It does not contradict our $\sech^2 x$ condition \eqref{eq:sech2_cond}.
One finds the relationship between the derivatives,
\begin{equation}
    \dv[2]{}{\xi} = -\dv[]{}{W} -2W \dv[2]{}{W}, 
\end{equation}
resulting in 
\begin{equation}
    \eval{\dv[2]{}{\xi}}_{\xi=0} = -\eval{\dv[]{}{W}}_{W=0}, 
\end{equation}
if the second energy derivatives of the distributions are finite for $W=0$.
Therefore, the $\sech^4 x$ potential condition given by Ref. \cite{Goldman2007} is equivalent to $C \ne 0$, which is consistent with our theory.  

The above derivations to obtain the $\sech^2 x$ potential \eqref{eq:phi} are the pseudo-potential method.
Another method to study the EHs is the BGK integral approach \cite{Bernstein1957,Turikov1984,Aravindakshan2018}.
Although the integral method can solve the accurate trapped distribution for the given potential and untrapped distribution, the pseudo-potential method has its own advantages.
The latter can give the general condition of the $\sech^2 x$ potential in the weak amplitude limit, i.e., Eq. \eqref{eq:sech2_cond}.
This condition does not rely on the specific form of $f_t$ and $f_u$.
However, 
the $\sech^2 x$ potential condition \eqref{eq:sech2_cond} cannot be derived by the integral approach.

\section{Applications to Kappa-distributed plasmas}
\label{sec:app}
\subsection{Theoretical results}
\label{sec:app_theo}
To specifically describe $\sech^2 x$ EHs, we consider the plasmas in which the undisturbed electrons follow the Kappa distribution.
The 1D Kappa velocity distribution is usually written as \cite{Mace1995}, 
\begin{equation}
    f_\kappa(v) = \frac{1}{\sqrt{\pi\kappa\theta^2}} \frac{\Gamma(\kappa)}{\Gamma(\kappa-1/2)} \left(1+\frac{v^2}{\kappa\theta^2}\right)^{-\kappa},
    \label{eq:kappa-vpdf}
\end{equation}
where $\theta$ is the most probable speed defined in the three-dimensional Kappa distribution \cite{Hellberg2009}.
The Kappa distribution recovers the Maxwellian one in the limit of $\kappa \rightarrow +\infty$,
so a less $\kappa$ parameter indicates a larger deviation from the Maxwellian equilibrium.
It should be noted that the $\kappa$ index must be larger than $3/2$ to ensure the convergence of the second moment of the three-dimensional Kappa distribution \cite{Livadiotis2009}.
The temperature is defined in a kinetic manner \cite{Livadiotis2009,Hellberg2009},
\begin{equation}
    \frac{1}{2}k_BT = \int_{-\infty}^{+\infty} \frac{1}{2}mv^2 f_\kappa(v) \dd{v}.
    \label{eq:T}
\end{equation}
leading to the parameter $\theta$ is,
\begin{equation}
    \theta = \sqrt{\frac{2\kappa-3}{\kappa}\frac{k_BT}{m}}.
    \label{eq:theta}
\end{equation}
In the literature, the Kappa distribution has two types of parameterizations,
i.e., whether the temperature $T$ or the most probable speed $\theta$ is independent of $\kappa$ \cite{Hellberg2009,Livadiotis2015a,Lazar2016}.
Lazar \textit{et al.} \cite{Lazar2016} suggested that $\theta$ should be independent of the kappa indices if the Kappa distribution is formed in some particle acceleration processes.
However, Yoon's work \cite{Yoon2014} implied that the temperature $T$ is constant if the Kappa distribution is generated due to the weak turbulence.
In addition, Refs. \cite{Hellberg2009} and \cite{Livadiotis2015a} claimed that the temperature should be an independent parameter from the consideration of statistical mechanics.
These two parameterizations could be valid but for different formations of Kappa distributions.
In this paper, we adopt the temperature $T$ as a $\kappa$-independent parameter.
The effects of another parameterization would be studied in the future.

In the solitary wave frame, the untrapped electrons are supposed to follow the shifted Kappa distribution,
\begin{equation}
    f_u^{\pm}(W) = N_\kappa \left[1+\frac{(\pm\sqrt{2W}+u)^2}{\kappa\theta^2}\right]^{-\kappa},
    \label{eq:fu}
\end{equation}
where $N_\kappa$ is the normalization factor,
\begin{equation}
    N_\kappa = \frac{1}{\sqrt{\pi\kappa\theta^2}} \frac{\Gamma(\kappa)}{\Gamma(\kappa-1/2)},
    \label{eq:Nk}
\end{equation}
and $u$ is the EH speed. 
The symbol $\pm$ in the distribution \eqref{eq:fu} takes plus for the right passing electrons and minus for the left passing ones.
To maintain consistency with Sec. \ref{sec:theo}, Eq. \eqref{eq:fu} should be expressed in the dimensionless form.
As mentioned in the first paragraph of Sec. \ref{sec:theo}, the energy $W$ is normalized by the unit $k_BT$ and the speed $u$ by $\sqrt{k_BT/m}$, so the parameter $\theta$ should be scaled by $\sqrt{k_BT/m}$.
Therefore, the dimensionless $\theta$ in Eq. \eqref{eq:fu} should be $\sqrt{(2\kappa-3)/\kappa}$ .

To obtain the $\sech^2 x$ EHs, the trapped electron distribution should meet the conditions $A,C=0$, i.e.,
\begin{equation}
    f_t|_{\xi=0}=\eval{\frac{f_u^++f_u^-}{2}}_{\xi=0}
    \quad \text{and} \quad
    \eval{\dv[2]{f_t}{\xi}}_{\xi=0}=\eval{\dv[2]{}{\xi}\left(\frac{f_u^++f_u^-}{2}\right)}_{\xi=0}.
    \label{eq:ft_cond}
\end{equation}
Thus, a general expansion of the trapped electron distribution for small-amplitude $\sech^2 x$ EHs could be constructed as,
\begin{align}
    f_t =& \eval{\frac{f_u^++f_u^-}{2}}_{\xi=0} + \eval{\dv[]{f_t}{\xi}}_{\xi=0} \cdot \xi + \eval{\dv[2]{}{\xi}\left(\frac{f_u^++f_u^-}{2}\right)}_{\xi=0}\cdot\frac{\xi^2}{2} + \eval{\dv[3]{f_t}{\xi}}_{\xi=0} \cdot \frac{\xi^3}{6}, \notag \\
        =& N_\kappa \left(1+\frac{u^2}{\kappa\theta^2}\right)^{-\kappa} \left\{1+\frac{\kappa \xi^2}{\kappa\theta^2+u^2}\left[1-\frac{2(\kappa+1)u^2}{\kappa\theta^2+u^2}\right]\right\}+ \eval{\dv[]{f_t}{\xi}}_{\xi=0} \cdot \xi + \eval{\dv[3]{f_t}{\xi}}_{\xi=0} \cdot \frac{\xi^3}{6},
    \label{eq:ft}
\end{align}
where the higher-order terms $O(\xi^4)$ are neglected.
In the limit of $\kappa \rightarrow +\infty$, if the first and third derivatives of $f_t$ at $\xi=0$ equal to zero,
Eqs. \eqref{eq:fu} and \eqref{eq:ft} reduce to,
\begin{equation}
    f_u^\pm = \frac{1}{\sqrt{2\pi}} e^{-\frac{1}{2}(\pm\sqrt{2W}+u)^2},
    \label{eq:fu_M}
\end{equation}
and
\begin{equation}
    f_t = \frac{1}{\sqrt{2\pi}} e^{-\frac{u^2}{2}} [1-(1-u^2)W].
    \label{eq:ft_M}
\end{equation}
In comparison with the original Schamel distribution \cite{Schamel1972,Korn1996}, Eq. \eqref{eq:fu_M} is the same as the untrapped Schamel distribution, while Eq. \eqref{eq:ft_M} is the first two terms of the expansion for the trapped Schamel distribution $(1/\sqrt{2\pi})\exp(-\beta W-u^2/2)$ with the condition $\beta=1-u^2$. 
It is mentioned in Ref. \cite{Korn1996} that such a requirement $\beta=1-u^2$ provides a $\sech^2 x$ solution in plasmas with the Schamel distribution.

After calculating the coefficients $B$ and $D$ for the distributions \eqref{eq:fu} and \eqref{eq:ft} (details in Appendix \ref{sec:dev_BDEF}), we derive the amplitude of the potential from Eqs. \eqref{eq:psi},
\begin{equation}
    \psi = -6 \frac{
            \pi\eval{\dv[]{f_t}{\xi}}_{\xi=0} - \frac{1}{\theta^2}\Re\left[U'_\kappa\left(\frac{u}{\theta}\right)\right]
        }
        {
            \pi\eval{\dv[3]{f_t}{\xi}}_{\xi=0} + \frac{1}{\theta^4}\Re\left[U'''_\kappa\left(\frac{u}{\theta}\right)\right]
        },
    \label{eq:psi_kappa}
\end{equation}
and the width from Eq. \eqref{eq:delta},
\begin{equation}
    \Delta^{-1} = \frac{1}{2}\sqrt{
            \pi\eval{\dv[]{f_t}{\xi}}_{\xi=0}
            -\frac{1}{\theta^2}\Re\left[U'_\kappa\left(\frac{u}{\theta}\right)\right]
    }, 
    \label{eq:delta_kappa}
\end{equation}
where the notation $\Re$ denotes the real part of the function and the apostrophe stands for the derivative.
$U_\kappa(\zeta)$ is the generalized dispersion function defined by \cite{Mace2009},
\begin{equation}
    U_\kappa(\zeta) = \frac{\Gamma(\kappa)}{\sqrt{\pi\kappa} \Gamma(\kappa-1/2)} \int_{-\infty}^{+\infty} \frac{(1+s^2/\kappa)^{-\kappa}}{s-\zeta} \dd{s},
    \label{eq:U}
\end{equation}
which could be numerically calculated from its hypergeometric function representation \cite{Mace2009},
\begin{equation}
    U_\kappa(\zeta) = i \frac{\kappa-1/2}{\kappa^{3/2}} {}_2F_1\left[1,2\kappa;\kappa+1;\frac{1}{2}\left(1-\frac{\zeta}{i\sqrt{\kappa}}\right)\right].
    \label{eq:U_2f1}
\end{equation}
The above derivations demonstrate that if the untrapped electron distribution is given, the profile of EHs is determined by the trapped electron distribution.
The continuities of the distribution and its second derivative at the separatrix, namely the conditions \eqref{eq:ft_cond}, ensure the $\sech^2 x$ form of EHs.
The first and third derivatives of $f_t$ at $\xi=0$ determine the amplitude \eqref{eq:psi_kappa} and the width \eqref{eq:delta_kappa} of the solitary-wave potential.

From Eqs. \eqref{eq:psi_kappa} and \eqref{eq:delta_kappa}, it seems that the EH potential with any $\psi$, $\Delta$, and $u$ could be constructed if a suitable $f_t$ is chosen.
However, some of them are not allowed because the constructed $f_t$ should be nonnegative.
The minimum of $f_t$ is taken at the peak of the potential $\phi=\psi$ and $v=0$, namely $\xi=\sqrt{2\psi}$ or equivalently $W=-\psi$. 
Therefore, $f_t(\xi=\sqrt{2\psi}) \ge 0$ is the condition restricting the scale of the amplitude and width.
\begin{figure}[ht]
	\centering
    \includegraphics[width=0.585\textwidth]{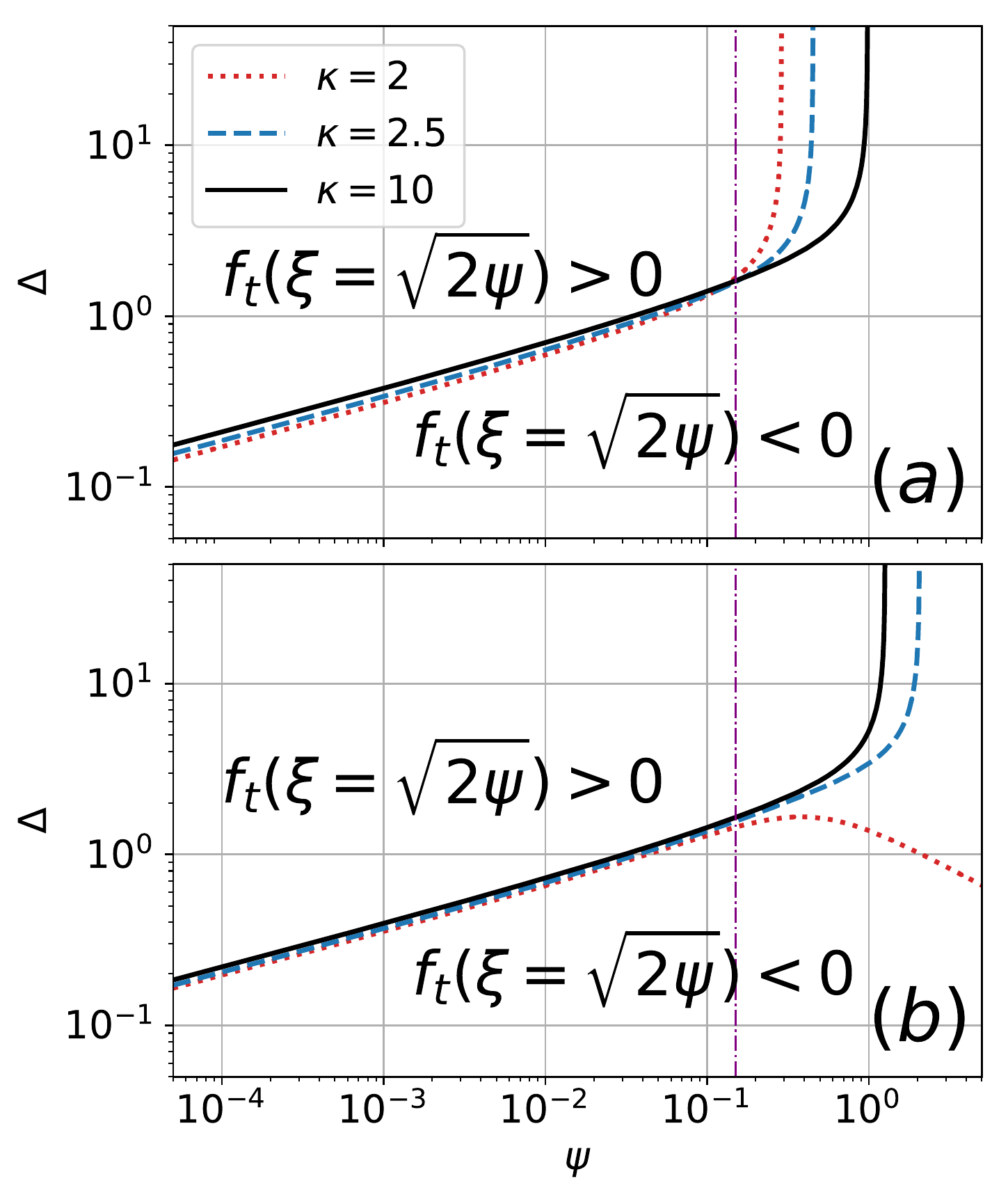}
    \caption{The parameter spaces of $\psi$ and $\Delta$ for different EH speeds (a) $u=0.1$, and (b) $u=0.4$. 
    The purple dash-dotted lines split the parameter space into two region, i.e., $\psi<0.15$ and $\psi \ge 0.15$.
        }
    \label{fig:parameter_spaces}
\end{figure}
\begin{figure}[ht]
	\centering
    \includegraphics[width=0.585\textwidth]{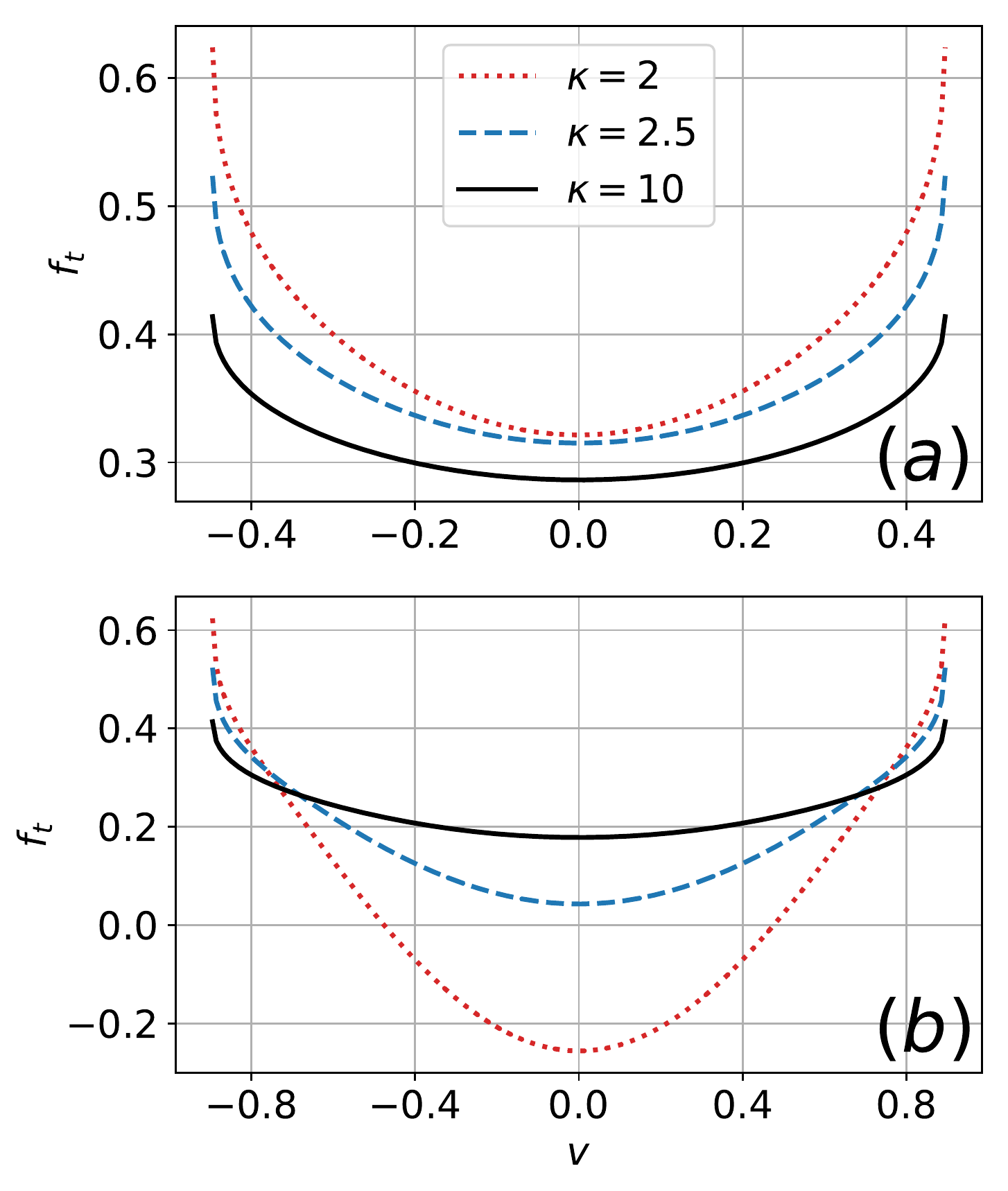}
    \caption{The velocity distribution of the trapped electrons at the peak of the potential $f_t|_{\phi=\psi}(v)$ for different kappa values in the case of (a) $\psi=0.1$ and (b) $\psi=0.4$.
    The potential width $\Delta=10$ and the EH speed $u=0.1$ are set for both (a) and (b).
        }
    \label{fig:ft}
\end{figure}

The allowable parameter spaces of $\psi$ and $\Delta$ are drawn in Fig. \ref{fig:parameter_spaces} for diverse EH speeds and kappa indices.
It should be mentioned that all the variables appeared in the figures are dimensionless throughout this paper.
These normalized variables are defined in the first paragraph of Sec. \ref{sec:theo}.
The value of $f_t(\xi=\sqrt{2\psi})$ is calculated from Eq. \eqref{eq:ft}. 
In this calculation, the first and third derivatives of $f_t$ at $\xi=0$ are solved with the given $\psi$ and $\Delta$ from Eqs. \eqref{eq:psi_kappa} and \eqref{eq:delta_kappa}.
For convenience, the parameter space is divided into two regions in Fig. \ref{fig:parameter_spaces}, i.e., $\psi<0.15$ and $\psi \ge 0.15$.

On the one hand, for the case of $\psi<0.15$, the potential width $\Delta$ has a lower limit.
This lower limit of $\Delta$ is slightly changed for varied EH speeds and kappa indices.
It shows in Fig. \ref{fig:parameter_spaces}(a) that the width limit could be lower for a smaller kappa index when the potential is weak.
This point could be explained as follows.
Because we compare the different Kappa distributions with the same temperature, the distribution with a small $\kappa$ would have more low-speed particles than that with a large $\kappa$.
As we know, the reduction of the electron density would support a positive potential.
A decreased potential width requires a larger reduction of the electron density.
Therefore, if there are more electrons at the EH speed, the reduction of them could support a more narrow potential.

On the other hand, for the case of $\psi \ge 0.15$, the limit of the potential amplitude $\psi$ is significantly affected by both EH speeds and kappa values.
From the perspective of EH speeds, the upper limit $\psi$ becomes large when the EH moves fast.
However, the amplitude limits that $\psi>1$ are beyond this work due to the weak potential assumption in our theory.
From the perspective of kappa indices, Fig. \ref{fig:parameter_spaces}(a) illustrates that the upper limit of $\psi$ decreases when $\kappa$ decreases for the slow EHs.
Compared with the case of $\psi<0.15$, more high-speed electrons would be trapped when $\psi \ge 0.15$.
Their behaviors can be described by the trapped electron distributions illustrated in Fig. \ref{fig:ft}(a).
It implies that a more reduction of trapped electron density is required for a less $\kappa$ to support the same potential.
The reason is that the electron density is contributed by both trapped and untrapped species.
To maintain the same potential, the total electron density must be the same for different kappa indices.
A less kappa value indicates more superthermal electrons, resulting in more untrapped particles passing through the solitary potential region.
Therefore, fewer electrons would be trapped.
During the increment of $\psi$, the changes from Fig. \ref{fig:ft}(a) to \ref{fig:ft}(b) show that $f_t(\xi=\sqrt{2\psi})$ with a less kappa index firstly reduces to a negative value for a slow EH with $u=0.1$.
Because the electron density cannot less than zero, the amplitude first reaches its maximum limit in the case of a small kappa parameter.
Therefore, the potential amplitude has a smaller upper limit with a smaller $\kappa$.

\subsection{Numerical self-consistent solutions}
\label{sec:app_num}

\begin{figure}[ht]
	\centering
    \includegraphics[width=0.5\textwidth]{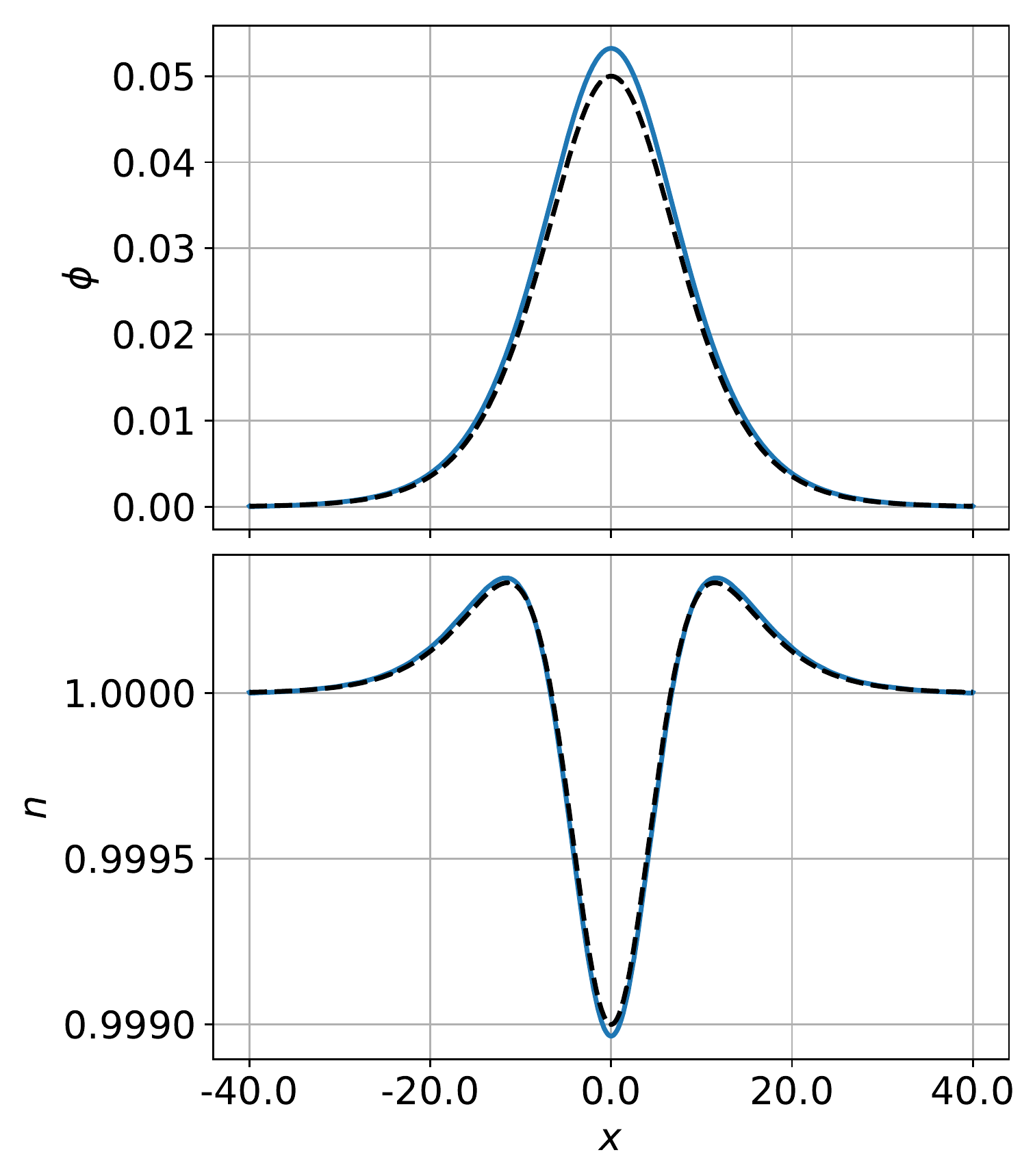}
    \caption{The potential (upper panel) and the electron number density (lower panel) solved from the self-consistent Poisson equation \eqref{eq:Poisson_sc} for $\psi=0.05$, $\Delta=10$, $u=1.0$ and $\kappa=5$.
            The numerical solutions are denoted by the blue solid curves, while the theoretical results by the black dashed curves.
    }
    \label{fig:verification_psi_n}
\end{figure}
\begin{figure}[ht]
	\centering
    \includegraphics[width=0.5\textwidth]{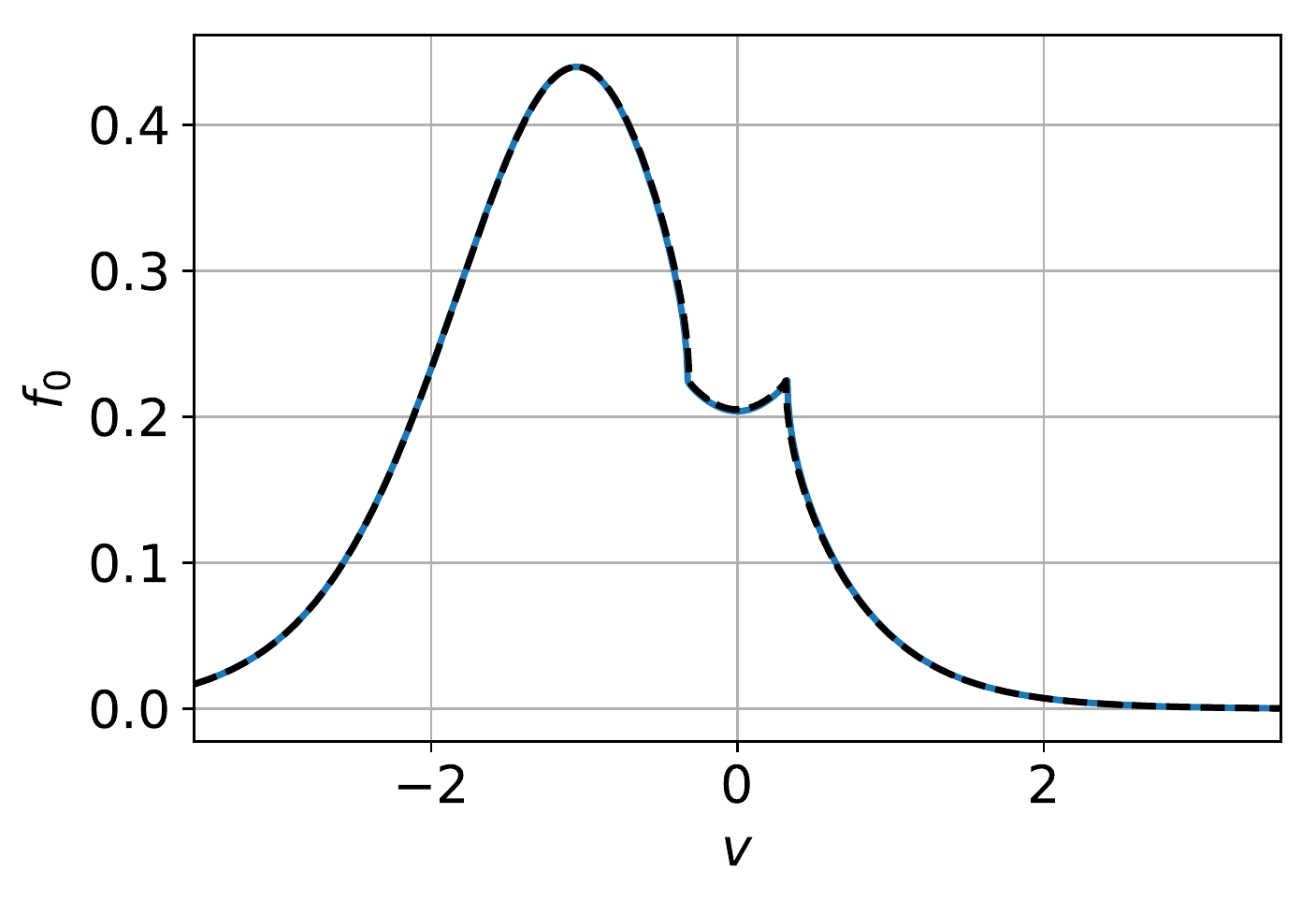}
    \caption{The velocity distribution at the peak of the potential $\phi=\psi$. The parameters, as well as the legends, are the same with those in Fig. \eqref{fig:verification_psi_n}. 
    }
    \label{fig:verification_fv}
\end{figure}
\begin{figure}[ht]
	\centering
    \includegraphics[width=0.5\textwidth]{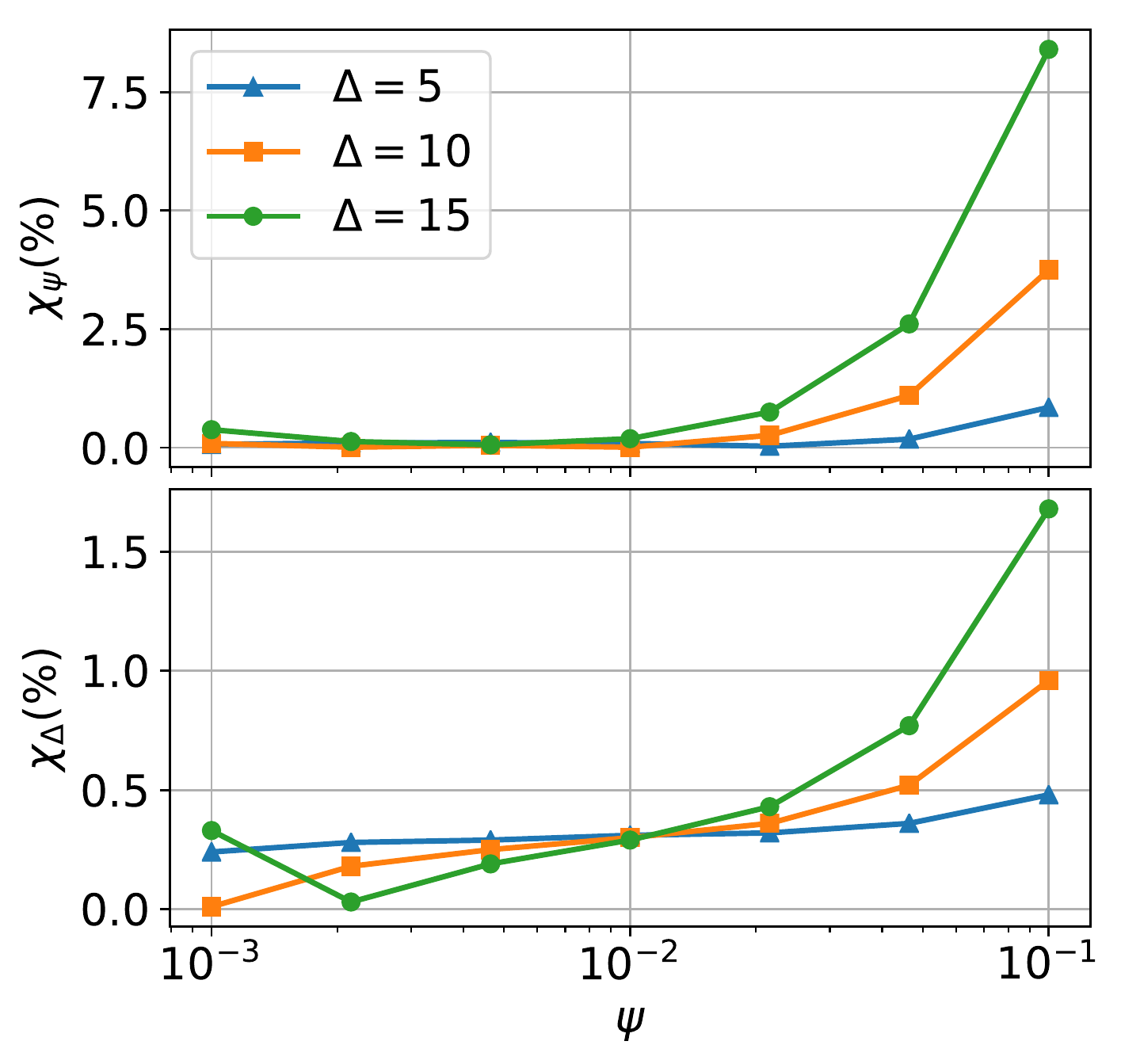}
    \caption{The relative errors of the amplitude (upper panel) and the width (lower panel) between the theoretical and numerical solutions in the case of $\kappa=5$ and $u=0.5$.
    We examine the theoretical results with the amplitude $\psi \in (0.001,0.1)$ and the width $\Delta=5$, $10$, $15$.
    The numerical amplitude and width are solved by fitting the numerical potential with the function $\psi_{num} \sech^2 (x/\Delta_{num})$.
    }
    \label{fig:verification_err1}
\end{figure}
\begin{figure}[ht]
	\centering
    \includegraphics[width=0.5\textwidth]{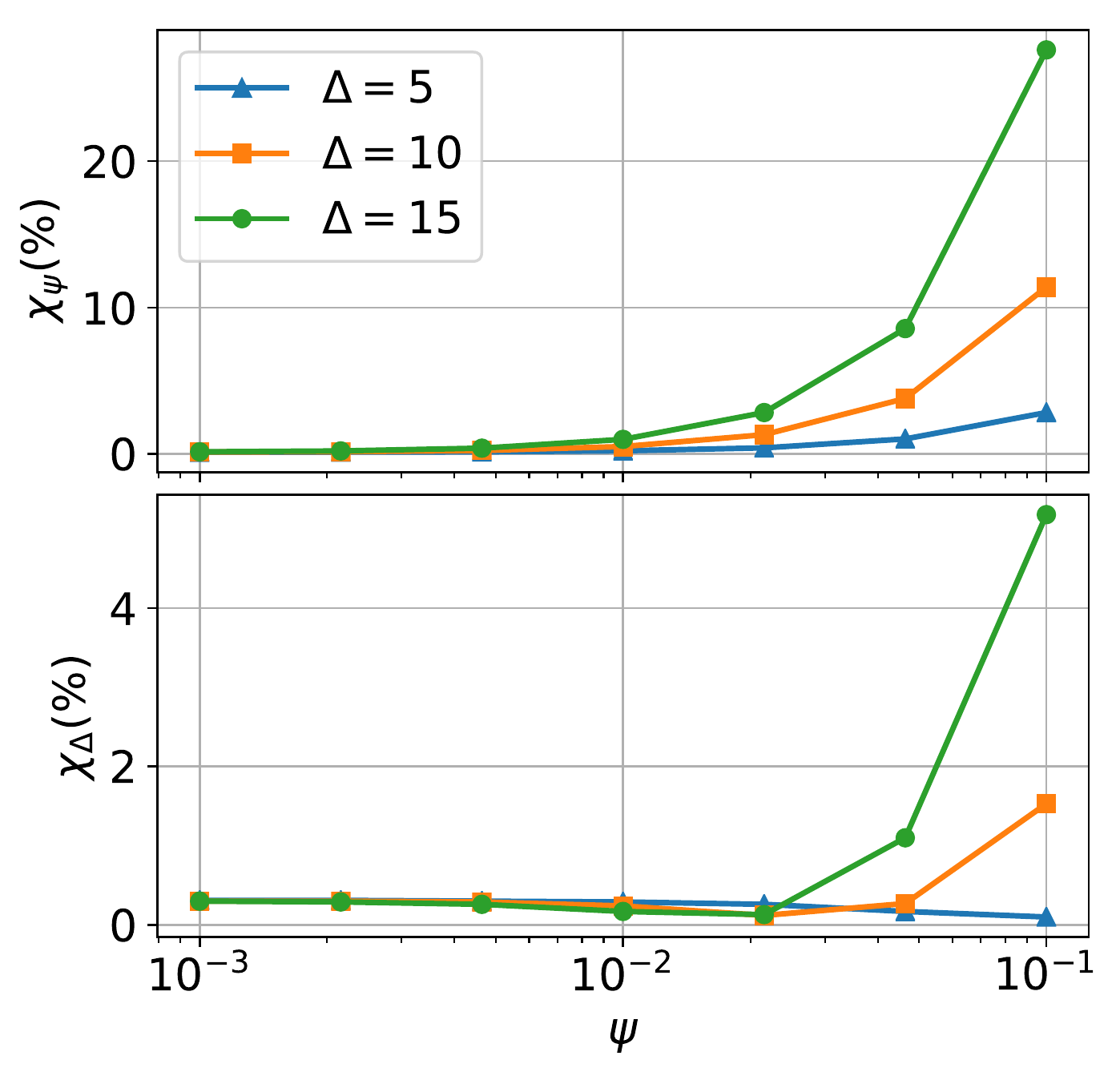}
    \caption{The relative errors of the amplitude (upper panel) and the width (lower panel) between the theoretical and numerical solutions in the case of $\kappa=8$ and $u=1.0$.
    }
    \label{fig:verification_err2}
\end{figure}
We conduct the numerical calculations to verify our theory.
Since the theory is valid for any amplitude $\psi \ll 1$ and width $\Delta$ satisfying $f_t(\xi=\sqrt{2\psi}) \ge 0$, we treat $\psi$ and $\Delta$ as free parameters.
Therefore, the theoretical potential refers to Eq. \eqref{eq:phi} with the predetermined $\psi$ and $\Delta$.
The numerical potential is obtained directly from the Poisson equation,
\begin{equation}
    \dv[2]{\phi}{x} = \int_{-\infty}^{+\infty} f\left(\frac{v^2}{2}-\phi\right) \dd{v} - 1,
    \label{eq:Poisson_sc}
\end{equation}
where the distribution $f$ is $f_u^\pm$ \eqref{eq:fu} for the untrapped electrons and $f_t$ \eqref{eq:ft} for the trapped electrons.
To construct the suitable trapped distribution $f_t$ for the given $\psi$ and $\Delta$,
we need the first and third derivatives of $f_t$ at $\xi=0$ in Eq. \eqref{eq:ft}.
These two derivatives can be solved from Eqs. \eqref{eq:psi_kappa} and \eqref{eq:delta_kappa} if $\psi$ and $\Delta$ are known.
With the determined distributions $f_u$ and $f_t$, the integral on the right side of Eq. \eqref{eq:Poisson_sc} is a function of $\phi$.
Therefore, this equation is a nonlinear ordinary differential equation of $\phi$.
After approximating the second derivative with the central difference, Eq. \eqref{eq:Poisson_sc} turns out to be a system of nonlinear equations which could be solved numerically by the Newton-Raphson algorithm \cite{Press2007}.
The potential is solved with the Dirichlet boundary conditions that $\phi$ is zero at both sides far away from the center.

The comparisons between theoretical and numerical solutions for the case of $\psi=0.05$, $\Delta=10$, $u=1.0$ and $\kappa=5$ are plotted in Figs. \ref{fig:verification_psi_n} and \ref{fig:verification_fv}.
The theoretical results are in good agreement with the numerical ones.
In addition, we verify the theoretical results with the numerical calculations on a large parameter scale.
By selecting different $\psi$ and $\Delta$, we obtain the corresponding numerical solutions.
The relative errors between theoretical and numerical results are plotted in Figs. \ref{fig:verification_err1} and \ref{fig:verification_err2}.
It indicates that the theory is more accurate with a smaller amplitude as expected.

\section{Summary}
\label{sec:sum}
The present work studies the kinetic theory of $\sech^2 x$ EHs by the pseudo-potential method.
We prove that if the electron distribution function satisfies the criteria \eqref{eq:sech2_cond}, the potential should be of form $\sech^2 x$, i.e., Eq. \eqref{eq:phi}, in the weak amplitude limit.
The amplitude \eqref{eq:psi} and the width \eqref{eq:delta} characterize the profile of the potential.
% Unlike Schamel's $\sech^4 x$ EHs, the trapped and untrapped distributions could be smooth at the separatrix for the $\sech^2 x$ EHs.

The theory is applied to Kappa-distributed plasmas.
The untrapped and trapped electron distributions are assumed to be \eqref{eq:fu} and \eqref{eq:ft}.
It is found that the potential amplitude, width, and speed are limited due to the nonnegativity of the trapped electron distribution $f_t$.
The parameter spaces of the amplitude and width are illustrated in Fig. \ref{fig:parameter_spaces}.
We find that the limit of the amplitude is significantly affected by EH speeds and kappa indices, but the width limit is only slightly changed.
The amplitude has an upper limit in the case of a slow $\sech^2 x$ EH but no limit for a fast one.
In addition, the width has an almost unchanged lower limit for different $u$ and $\kappa$.
To verify our theory, we numerically calculate the self-consistent solution of the Poisson equation \eqref{eq:Poisson_sc}.
The results are shown in Figs. \ref{fig:verification_psi_n}-\ref{fig:verification_err2}.
It demonstrates that the relative errors between the numerical and theoretical results decrease as the potential tends to the weak amplitude limit.

In the literature \cite{Goldman2007,Schamel2015,Zhou2016,Hutchinson2017}, the solitary potential of EHs is always considered to be of the $\sech^4 x$ form.
Goldman \textit{et al.} \cite{Goldman2007} showed that the $\sech^4 x$ potential is unique in the weak amplitude limit if the energy derivatives of the trapped and untrapped distributions are unequal at the separatrix.
However, the present work provides an alternative possibility that the solitary potential could be of the $\sech^2 x$ form for small-amplitude EHs.
Such a potential shape might be used in the observation analyses and simulations.
Besides, we first give the general kinetic theory of the $\sech^2 x$ EHs and analyze their physical properties in Kappa-distributed plasmas.
It will improve our understanding of the characteristics of EHs with diverse potential shapes in non-thermal plasmas. 

\begin{acknowledgments}
This work was supported by the National Natural Science Foundation of China (No.12105361) and by the Supporting Fund from Civil Aviation University of China (No.3122022PT18).
\end{acknowledgments}

\section*{Data Availability}
No new data were created or analysed in this study.

\appendix
\section{The derivations for the expansions of $I_1$ and $I_2$}
\label{sec:dev_I1}
We expand $I_1$, namely Eq. \eqref{eq:n_phi_i1}, into a power series at $\phi =0$,
\begin{equation}
    I_1 =\sum_{k=0}^\infty \left[\frac{(-1)^k}{k!} \int_{-\infty}^{+\infty} \eval{\dv[k]{f_u}{W}}_{\phi=0} \dd{v}\right] \cdot \phi^k. 
    \label{ap:I1_series_1}
\end{equation}
It is same that taking $\phi=0$ after or before the $k$-order derivatives of $f_u$,
so we have,
\begin{equation}
    \eval{\dv[k]{f_u}{W}}_{\phi=0}=\dv[k]{F_u}{W}=\left(\frac{1}{v}\dv{}{v}\right)^k F_u,
    \label{ap:dev_relation}
\end{equation}
where $F_u = f_u|_{\phi=0}$ is the free electron distribution for $\phi=0$, or equivalently $x \rightarrow \pm \infty$.
With the aid of Eq. \eqref{ap:dev_relation}, the integral in Eq. \eqref{ap:I1_series_1} could be integrated by parts,
\begin{align}
    \int_{-\infty}^{+\infty} \eval{\dv[k]{f_u}{W}}_{\phi=0}\dd{v}
        =& \pv{\int_{-\infty}^{+\infty} \left(\frac{1}{v}\dv{}{v}\right)^k F_u \dd{v}} \notag \\
        =& \pv{\int_{-\infty}^{+\infty} \frac{1}{v} \dd{\left[\left(\frac{1}{v}\dv{}{v}\right)^{k-1} F_u\right]}} \notag \\
        =& \pv{\int_{-\infty}^{+\infty} \frac{1}{v^2} \left(\frac{1}{v}\dv{}{v}\right)^{k-1} F_u \dd{v}},
\end{align}
where the symbol $\pv$ stands for the Cauchy principal value of the integral. 
In the integrations by parts, we assume $\frac{1}{v} \left(\frac{1}{v}\dv{}{v}\right)^{k-1} F_u$ vanishes at the boundary $v\rightarrow \pm \infty$.
After repeating the integrations by parts for $k$ times, $I_1$ is reduced to,
\begin{equation}
    I_1 =1+\sum_{k=1}^\infty \left[\frac{(-1)^k(2k-1)!!}{k!} \pv{\int_{-\infty}^{+\infty} \frac{F_u}{v^{2k}} \dd{v}}\right] \cdot \phi^k, 
    \label{ap:I1_series}
\end{equation}
where the normalization condition $\int_{-\infty}^{+\infty} F_u \dd{v}=1$ is used.

The expansion of $I_2$ can be written as,
\begin{align}
    I_2 =&\int_{-\sqrt{2\phi}}^{\sqrt{2\phi}} f_t \dd{v} - \int_0^{\sqrt{2\phi}} f_u^+ \dd{v} - \int_{-\sqrt{2\phi}}^0 f_u^- \dd{v} \notag \\
        =&\sum_{k=0}^\infty \frac{1}{k!} \left(
            \eval{\dv[k]{f_t}{\xi}}_{\xi=0} \int_{-\sqrt{2\phi}}^{\sqrt{2\phi}} \xi^k \dd{v}
            -\eval{\dv[k]{f_u^+}{\xi}}_{\xi=0} \int_{0}^{\sqrt{2\phi}} \xi^k \dd{v}
            -\eval{\dv[k]{f_u^-}{\xi}}_{\xi=0} \int_{-\sqrt{2\phi}}^{0} \xi^k \dd{v}
            \right). 
    \label{ap:I2_dev}
\end{align}
Since $\xi = \sqrt{2\phi-v^2}$ is an even function of $v$, one has $\int_{0}^{\sqrt{2\phi}} \xi^k \dd{v} = \int_{-\sqrt{2\phi}}^{0} \xi^k \dd{v} = \frac{1}{2} \int_{-\sqrt{2\phi}}^{\sqrt{2\phi}} \xi^k \dd{v}$.
Thus, Eq. \eqref{ap:I2_dev} becomes,
\begin{equation}
    I_2 =\sum_{k=0}^\infty \frac{1}{k!} \eval{\dv[k]{}{\xi} \left( f_t - \frac{f_u^++f_u^-}{2}\right)}_{\xi=0}
             \cdot \int_{-\sqrt{2\phi}}^{\sqrt{2\phi}} \xi^k \dd{v}. 
    \label{ap:I2_dev2}
\end{equation}
We directly calculate the integral,
\begin{equation}
    \int_{-\sqrt{2\phi}}^{\sqrt{2\phi}} \xi^k \dd{v} = \frac{\sqrt{\pi}\Gamma\left(\frac{k+2}{2}\right)}{\Gamma\left(\frac{k+3}{2}\right)} (2\phi)^{\frac{k+1}{2}}.
    \label{ap:integral_xi}
\end{equation}
After taking Eq. \eqref{ap:integral_xi} back to Eq. \eqref{ap:I2_dev2}, one gets the expansion of $I_2$ \eqref{eq:I2_series}.

\section{The derivations of the $\sech^2 x$ potential}
\label{sec:dev_phi}
The electron number density is reduced to,
\begin{equation}
    n(\phi) = 1 + B\phi + D\phi^2,
    \label{ap:n}
\end{equation}
in the case of $A,C=0$ and $B,D \neq 0$ by neglecting the higher-order terms $O(\phi^{5/2})$ in Eq. \eqref{eq:n_abcd}.
According to the pseudo-potential approach \cite{Hutchinson2017}, the Sagdeev potential is obtained by,
\begin{equation}
    V(\phi) =\int_0^\phi (1-n) \dd{\phi} = -\frac{1}{2}B\phi^2-\frac{1}{3}D\phi^3.
    \label{eq:Sagdeev_potential}
\end{equation}
The nonlinear dispersion relation $V(\psi)=0$ leads to,
\begin{equation}
    \psi = -\frac{3B}{2D}.
\end{equation}
Then, the potential is implicitly solved from the Poisson equation \eqref{eq:Poisson},
\begin{equation}
    x = \int_\phi^\psi \frac{\dd{\phi}}{\sqrt{-2V(\phi)}} = \frac{1}{\sqrt{B}} \ln(\frac{1+\sqrt{1-\frac{\phi}{\psi}}}{1-\sqrt{1-\frac{\phi}{\psi}}}),
    \label{ap:phi_eq}
\end{equation}
for $\psi>0$.
Exponentiating the both sides of Eq. \eqref{ap:phi_eq}, one has,
\begin{equation}
    e^{\sqrt{B}x} = \frac{1+\sqrt{1-\frac{\phi}{\psi}}}{1-\sqrt{1-\frac{\phi}{\psi}}} \quad \text{and} \quad
    e^{-\sqrt{B}x} = \frac{1-\sqrt{1-\frac{\phi}{\psi}}}{1+\sqrt{1-\frac{\phi}{\psi}}},
\end{equation}
resulting in,
\begin{equation}
    \cosh(\sqrt{B}x) = \frac{e^{\sqrt{B}x}+e^{-\sqrt{B}x}}{2} = 2\frac{\psi}{\phi}-1.
    \label{ap:cosh}
\end{equation}
Eventually, the potential is explicitly expressed as,
\begin{equation}
    \phi(x) = \psi \sech[2](\frac{\sqrt{B}x}{2}).
\end{equation}

\section{The calculations of the expansion coefficients in Kappa-distributed plasmas}
\label{sec:dev_BDEF}
In Kappa-distributed plasmas, we assume the untrapped electron distribution is $f_u$ \eqref{eq:fu} and the trapped electron distribution is $f_t$ \eqref{eq:ft}.
In terms of the definitions \eqref{eq:B} and \eqref{eq:D}, one finds,
\begin{equation}
    B = \pi \eval{\dv[]{f_t}{\xi}}_{\xi=0} - N_\kappa\pv{\int_{-\infty}^{+\infty} \frac{\left[1+\frac{(v+u)^2}{\kappa\theta^2}\right]^{-\kappa}}{v^2}\dd{v}},
    \label{ap:B_dev}
\end{equation}
and
\begin{equation}
    D = \frac{\pi}{4} \eval{\dv[3]{f_t}{\xi}}_{\xi=0} + \frac{3}{2}N_\kappa\pv{\int_{-\infty}^{+\infty} \frac{\left[1+\frac{(v+u)^2}{\kappa\theta^2}\right]^{-\kappa}}{v^4}\dd{v}}.
    \label{ap:D_dev}
\end{equation}
We can calculate a more general integral,
\begin{align}
    &N_\kappa \pv{\int_{-\infty}^{+\infty} \frac{\left[1+\frac{(v+u)^2}{\kappa\theta^2}\right]^{-\kappa}}{v^{2k}}\dd{v}} \notag \\
    =& \frac{1}{\theta^{2k}}\frac{\Gamma(\kappa)}{\sqrt{\pi\kappa} \Gamma(\kappa-1/2)} \pv{\int_{-\infty}^{+\infty} \frac{(1+t^2/\kappa)^{-\kappa}}{(t-u/\theta)^{2k}} \dd{t}}. \notag \\
    =& \frac{1}{(2k-1)!\theta^{2k}} \Re \left[U_\kappa^{(2k-1)} \left(\frac{u}{\theta}\right)\right],
    \label{ap:general_integral}
\end{align}
where the substitution $t=(v+u)/\theta$ is used.
$U_\kappa^{(2k-1)}(\zeta)$ is the $(2k-1)$-order derivative of $U_\kappa(\zeta)$ defined in Eq. \eqref{eq:U}.
Thus, with the help of Eq. \eqref{ap:general_integral}, the coefficients $B$ and $D$ are derived,
\begin{equation}
    B = \pi \eval{\dv[]{f_t}{\xi}}_{\xi=0} - \frac{1}{\theta^2} \Re\left[U'_\kappa\left(\frac{u}{\theta}\right)\right],
    \label{ap:B_kappa}
\end{equation}
and
\begin{equation}
    D = \frac{\pi}{4} \eval{\dv[3]{f_t}{\xi}}_{\xi=0} + \frac{1}{4\theta^4} \Re\left[U'''_\kappa\left(\frac{u}{\theta}\right)\right].
    \label{ap:D_kappa}
\end{equation}
The generalized dispersion function $U_\kappa(\zeta)$ \eqref{eq:U} recovers the standard dispersion function $Z(\zeta)$ when $\kappa$ goes to infinity \cite{Mace2009}. 
Therefore, if the derivatives of $f_t$ at $\xi=0$ vanish,
\begin{equation}
    \eval{\dv[]{f_t}{\xi}}_{\xi=0}=\eval{\dv[3]{f_t}{\xi}}_{\xi=0}=0,
\end{equation}
the coefficients $B$ and $D$ in the limit $\kappa\rightarrow \infty$ turn to be,
\begin{equation}
    B = -\frac{1}{2}\Re\left[Z'\left(\frac{u}{\sqrt{2}}\right)\right] \quad \text{and} \quad 
    D = \frac{1}{16} \Re\left[Z'''\left(\frac{u}{\sqrt{2}}\right)\right],
\end{equation}
which are consistent with the expansion of the electron number density $n$ derived in Ref. \cite{Korn1996}.

\bibliography{mylib}
\end{document}